\documentclass[fleqn]{article}

\usepackage{amsmath,graphicx}

\begin{document}

\title{Relativistic corrections of order $m\alpha^6$ to the two-center problem}

\author{V.I.~Korobov and Ts.~Tsogbayar\\
\it Joint Institute for Nuclear Research, 141980, Dubna,
Russia}
\date{}

\maketitle

\begin{abstract}
Effective potentials of the relativistic $m\alpha^6$ order correction for
the ground state of the Coulomb two-center problem are calculated. They
can be used to evaluate the relativistic contribution of that order
to the energies of hydrogen molecular ions or metastable states of the
antiprotonic helium atom, where precision spectroscopic data are
available. In our studies we use the variational expansion based on
randomly chosen exponents that permits to achieve high numerical accuracy.
\end{abstract}

\section{Introduction}
In recent years several experiments on precision spectroscopy of
three-body light atomic and molecular systems become available
\cite{Hori06,Sch06,Koe07}. That is a great challenge to theorists, since
the quantum electrodynamics for the few-body bound state problem is not
that well elaborated as for atoms comprised of two particles.

On the other hand, numerical variational solution for the nonrelativistic
Coulomb three--body problem have reached very high precision. The ground
state of the $\mbox{H}^+_2$ molecular ion has been calculated with as much
as 30 digits \cite{Yan07} (for other examples of nonrelativistic
calculations see references therein).

Accurate knowledge of both theoretical and experimental transition
frequencies has the potential of improving $m_e/m_p$, $m_e/m_{\bar{p}}$,
etc mass ratios \cite{Gre98,Sch03,Yam02}. In order to achieve this goal
the important step is evaluating the relativistic and radiative
corrections for the binding energies. This can be systematically performed
using series expansion of the binding energy in terms of the coupling
constant, $\alpha$. The leading order relativistic and radiative
corrections ($R_\infty\alpha^2$, $R_\infty\alpha^3$) have been obtained in
\cite{Kor03,Kor06}. The main aim of this work is to calculate a
contribution of order $m\alpha^6$ (or $R_\infty \alpha^4$) due to
relativistic corrections for the bound electron in the nonrecoil limit.
The radiative corrections to this order are known analytically
\cite{SapYen}. A sum of these two contributions constitutes a complete
$m\alpha^6$ order correction for a one electron three-body
system in a nonrecoil limit.

\section{Nonrelativistic Hamiltonian and Variational Expansion}

In what follows we consider the Coulomb two--center problem with the
nonrelativistic Hamiltonian
\begin{equation}\label{Hamiltonian}
H_0 = \frac{\mathbf{p}^2}{2m}+V,
\qquad
V = -\frac{Z_1}{r_1}-\frac{Z_2}{r_2}\,.
\end{equation}
where $r_1$ and $r_2$ are the distances from an electron to nuclei 1 and
2, respectively. Atomic units, $m_e\!=\!e\!=\!\hbar\!=\!1$, are used
throughout.

In order to get a precise solution for the Schr\"odinger equation
\begin{equation}\label{nonrel}
\left[\frac{\mathbf{p}^2}{2m}+V\right]\Psi_0(\mathbf{r}) =
  E_0\Psi_0(\mathbf{r}),
\end{equation}
we use the variational approach. A variational expansion for the wave
function of the ground state of electron is taken in the form ($Z_1\ne Z_2$):
\begin{equation}\label{exp}
   \Psi(\mathbf{r}) = e^{im\varphi}r^{|m|}\sum^{\infty}_{i=1}
 C_{i}e^{-\alpha_{i} r_1 - \beta_{i} r_2},
\end{equation}
where $r$ is a distance from an electron to the $z$-axis and
\[
r=\frac{1}{2R}\sqrt{2r_1^2r_2^2+2r_1^2R^2+2r_2^2R^2-r_1^4-r_2^4-R^4}.
\]
For $Z_1=Z_2$ the variational wave function should be symmetrized
\begin{equation}\label{expsym}
   \Psi(\mathbf{r_{1},r_{2}}) = e^{im\varphi}r^{|m|}\sum^{\infty}_{i=1}
 C_{i}(e^{-\alpha_{i} r_{1} - \beta_{i} r_{2}}\pm
       e^{-\beta_{i} r_{1} - \alpha_{i} r_{2}} ),
\end{equation}
where $(+)$ is used to get a {\em gerade} electronic state and $(-)$ is
for an {\em ungerage} state, respectively. Parameters
$\alpha_{i}$ and $\beta_{i}$ are generated in a quasi-random
manner~\cite{Kor00,Bai02}
\begin{equation}\label{generator}
  \alpha_{i}=
  \left\lfloor\frac{1}{2}i(i+1)\sqrt{p_{\alpha}}\right\rfloor
                              (A_{2}-A_{1}) + A_{1}.
\end{equation}
Here $\lfloor x\rfloor$ designates the fractional part of $x$,
$p_{\alpha}$ is a prime number, an interval $[A_{1},A_{2}]$ is a real
variational interval, which has to be optimized. Parameters $\beta_{i}$
are obtained in a similar way.

The Schr\"odinger equation expressed in coordinates, $r_1$, $r_2$,
$\varphi$, has a form:
\begin{equation}
\begin{array}{@{}l}
\displaystyle
\biggl\{
   -\frac{1}{2}\left[
       \left(\partial_{r_1}^2\!+\!\frac{2}{r_1}\partial_{r_1}\right)
      +\left(\partial_{r_2}^2\!+\!\frac{2}{r_2}\partial_{r_2}\right)
      +\frac{r_1^2\!+\!r_2^2\!-\!R^2}{r_1r_2}\>\partial_{r_1}\partial_{r_2}
      -\frac{m^2}{r^2}
   \right]
\\[4mm]\displaystyle\hspace{60mm}
      -\frac{Z_1}{r_1}-\frac{Z_2}{r_2}
\biggr\}\Psi
   = E\Psi.
\end{array}
\end{equation}

In order to get accurate results we use several sets of basic functions of
Eq.~(\ref{exp}) (or (\ref{expsym})) (in a spirit of \cite{Kor00}) with
different values of variational parameters: three sets for small values of
internuclear distance $R$, and two sets for intermediate and large values
of $R$, respectively. Total number of basis functions varies from $N=110$
to $N=250$. Using the variational method described above the
nonrelativistic energy has been obtained with accuracy of more than 20
significant digits. In these calculations arithmetics of sextuple
precision (about 48 decimal digits) has been used. Earlier, in a previous
paper \cite{Tso06} using the same variational approach the effective
potentials of the Breit-Pauli Hamiltonian have been calculated with
up to 10 significant digits.

The matrix elements, which appear in this and following sections, have
been evaluated analytically, for further details see Appendix A.

\section{Breit--Pauli Hamiltonian}

The leading order relativistic correction of order $m\alpha^{4}$
can be expressed by the Breit-Pauli Hamiltonian:
\begin{equation}\label{BP}
\begin{array}{@{}l}
\displaystyle
H_{B} = -\frac{\mathbf{p}^4}{8m^3}
         +\frac{1}{8m^{2}}[Z_{1}4\pi\delta(\mathbf{r}_{1})
                          +Z_{2}4\pi\delta(\mathbf{r}_{2})]+
\\[2mm]\displaystyle\hspace{50mm}
      \left(
          Z_1\frac{[\mathbf{r}_1\times\mathbf{p}]}{2m^2r_1^3}+
          Z_2\frac{[\mathbf{r}_2\times\mathbf{p}]}{2m^2r_2^3}
      \right)\mathbf{s}\>,
\end{array}
\end{equation}

%In order to calculate expectation value of $\mathbf{p}^{4}$ we use the
%relation
%\[
% \mathbf{p}^{2}\Psi_0 = 2 m (E_{0}-V)\Psi_0,
%\]
%where $(E_{0}, \Psi_0)$ is a solution of the Schr\"odinger equation
%(\ref{nonrel}), that allows to get a less singular expression
%\begin{equation}\label{1.2}
%\left\langle \mathbf{p}^{4}\right\rangle =
%   \left\langle 4m^{2}(E_{0}-V)^{2}\right\rangle.
%\end{equation}

The spin-dependent term (the last term in Eq.~(\ref{BP})) of the
Breit-Pauli Hamiltonian $H_B^{so}$ does not contribute to the energy at
this order, but should be taken into account, when the $m\alpha^6$ order
relativistic corrections are considered.

\section{Relativistic corrections of $m\alpha^6$ order}

In this section we will assume for simplicity of notation that $H_B\equiv
H_B^s$, the scalar term of $H_B$. The spin-orbit term $H_B^{so}$ will be
treated separately.

Our method is based on the effective Hamiltonian approach, which is
closely related to ideas of \cite{Pac97,Adk05}, where a case of atomic
hydrogen has been considered.

\subsection{Formal expressions}

The energy displacement due to the $m\alpha^6$ order relativistic
corrections can be expressed as:
\begin{equation}\label{e6}
\Delta E^{(6)} =
   \left\langle H_B Q (E_0-H_0)^{-1} Q H_B \right\rangle
  +\left\langle H^{(6)} \right\rangle.
\end{equation}
Here $Q=I\!-|\psi_0\rangle\langle\psi_0|\,$ is a projection operator.
The effective Hamiltonian corresponding to this contribution has a form
\cite{Pac97}
\begin{equation}\label{h6}
\begin{array}{@{}l}
\displaystyle
H^{(6)} = \frac{p^6}{16m^5}
     +\frac{(\boldsymbol{\mathcal{E}}_1\!+\!\boldsymbol{\mathcal{E}}_2)^2}
                                                                {8m^3}
     -\frac{3\pi}{16m^4}
        \Bigl\{
           p^2\bigl[\rho_1\!+\!\rho_2\bigr]+
           \bigl[\rho_1\!+\!\rho_2\bigr]p^2
        \Bigr\}
\\[4mm]\displaystyle\hspace{44mm}
     +\frac{5}{128m^4}\left(p^4V\!+\!Vp^4\right)
     -\frac{5}{64m^4}\left(p^2Vp^2\right),
\end{array}
\end{equation}
where $\boldsymbol{\mathcal{E}}_i=-Z_i\mathbf{r}_i/r_i^3$ and
$\rho_i=Z_i\delta(\mathbf{r}_i)$ ($\Delta V=4\pi\rho$).

Separately, both contributions of (\ref{e6}) are divergent.

\subsection{Removing divergences from the second order contribution}

The second order perturbation term
\begin{equation}\label{HB2}
\Delta E^{\{2\}}_B = \left\langle H_B Q (E_0-H_0)^{-1} Q H_B \right\rangle
\end{equation}
can be evaluated by obtaining the wave function $\Psi_B$ as a solution
of equation
\begin{equation}
(E_0-H_0)\Psi_B = (H_B-\langle H_B \rangle)\Psi_0,
\end{equation}
then the correction to the energy from Eq.~(\ref{HB2}) can be expressed as
$\Delta E^{\{2\}}_B = \langle\Psi_0 |(H_B-\langle H_B \rangle)|\Psi_B\rangle$.

It is known that a formal second order expression of the type
\[
\left\langle
\Psi_0|\delta^3(\mathbf{r}_i)Q(E_0-H_0)^{-1}Q\delta^3(\mathbf{r}_i)|\Psi_0
\right\rangle
\]
is divergent. Similarly, $p^4|\Psi_0\rangle$ behaves as
\[
-2m p^2V|\Psi_0\rangle \sim 8\pi m[Z_1\delta(\mathbf{r}_1)
\!+\!Z_2\delta(\mathbf{r}_2)]|\Psi_0\rangle
\]
at small values of $\mathbf{r}_1$, $\mathbf{r}_2$.

In order to understand how to overcome this problem, let us consider first
the atomic hydrogen $S$-state case. Let $\Psi_B^{(H)}$ be a
solution of equation
\begin{equation}
\label{WFBreit}
\left[E_0-\left(\frac{p^2}{2m}-\frac{Z}{r}\right)\right]\Psi_B^{(H)}=
   Q\left[
      -\frac{p^4}{8m^3}+\frac{Z\pi}{2m^2}\delta(\mathbf{r})
   \right]\Psi_0^{(H)},
\end{equation}
separating the dominant behaviour at small $r$, the solution of above
equation can be presented as
\[
\Psi_B^{(H)}= \frac{Z\Psi_0(0)}{4mr}
%       -\frac{3Z^2\Psi_0(0)+2Z\Psi'_0(0)}{2}\ln{r}
       +\tilde{\Psi}_B^{(H)},
\]
where $\tilde{\Psi}_B^{(H)}$ is a less singular function,
$\tilde{\Psi}_B^{(H)}\sim\ln{r}$ at $r\to0$.

Coming back to the two center problem, let us try to separate the singular
part of the Breit-Pauli wave function solution $\Psi_B$ as follows:
\begin{equation}\label{PBmod}
\Psi_B = U\Psi_0 + \tilde{\Psi}_B.
\end{equation}
where $U=-\frac{1}{4m}V$. Substituting of Eq.~(\ref{PBmod}) into
Eq.~(\ref{HB2}) modifies our equations,
\begin{subequations}
\begin{equation}\label{eq:s1a}
\Delta E_B^{\{2\}}=
   \left\langle\Psi_0|(H_B-\langle H_B \rangle)U|\Psi_0\right\rangle
   +\left\langle
      \Psi_0|(H_B-\langle H_B\rangle)|\tilde{\Psi}_B
   \right\rangle,
\end{equation}
and $\tilde{\Psi}_B$ is a solution of
\begin{equation}\label{eq:s1b}
(E_0-H_0)\tilde{\Psi}_B =
   -(E_0-H_0)U\Psi_0+(H_B-\langle H_B\rangle)\Psi_0.
\end{equation}
\end{subequations}
However, the last term in Eq.~(\ref{eq:s1a}) is still divergent, the
singularity from the left-hand side of this term should be eliminated as
well:
\begin{equation}
\begin{array}{@{}l}
\left\langle
   \Psi_0|(H_B\!-\!\langle H_B \rangle)|\tilde{\Psi}_B
\right\rangle
\\[3mm]\hspace{10mm}
=  \bigl\langle
      \Psi_0|(H_B\!-\!\langle H_B \rangle)(E_0\!-\!H_0)^{-1}
\\[2mm]\hspace{35mm}\times
      [-(E_0\!-\!H_0)U+(H_B\!-\!\langle H_B\rangle)]|\Psi_0
   \bigr\rangle
\\[3mm]\hspace{10mm}
 = \left\langle
      \Psi_0|-U(E_0\!-\!H_0)U+U(H_B\!-\!\langle H_B\rangle)|\Psi_0
   \right\rangle
\\[1.5mm]\hspace{35mm}
 + \left\langle
      \tilde{\Psi}_B|-(E_0\!-\!H_0)U+(H_B\!-\!\langle H_B\rangle)|\Psi_0
   \right\rangle
\end{array}
\end{equation}
These transformations as can be seen are equivalent to the one used by
Pachucki in \cite{Pac00}:
\begin{equation}\label{transform}
\left\{
\begin{array}{@{}l}
H'_B = H_B-(E_0\!-\!H_0)U-U(E_0\!-\!H_0)\\[2mm]
\displaystyle
\left\langle H_B Q (E_0\!-\!H_0)^{-1} Q H_B \right\rangle =
     \left\langle H'_B Q (E_0\!-\!H_0)^{-1} Q H'_B \right\rangle
\\[1.5mm]\displaystyle\hspace{20mm}
    +\left\langle UH_B\!+\!H_BU \right\rangle
%\\[2mm]\displaystyle\hspace{50mm}
    -2\left\langle U \right\rangle \left\langle H_B \right\rangle
    -\left\langle U(E_0\!-\!H_0)U \right\rangle
\end{array}
\right.
\end{equation}
The last three terms can be recast in a form of a new interaction modifying
the Hamiltonian $H^{(6)}$:
\begin{equation}\label{H6p}
\begin{array}{@{}l}
\displaystyle
H'^{(6)} =
    (UH_B+H_BU)-2U\langle H_B \rangle-U(E_0-H_0)U,
\\[3mm]\displaystyle
\langle H'^{(6)}\rangle =
    \frac{1}{32m^4}\left\langle p^4V\!+\!Vp^4 \right\rangle
    -\frac{\pi}{4m^3}\left\langle\>
        V\!\left[\rho_1\!+\!\rho_2\right]\>
     \right\rangle
\\[2mm]\displaystyle\hspace{40mm}
    +\frac{1}{32m^3}\left\langle
        (\boldsymbol{\mathcal{E}}_1\!+\!\boldsymbol{\mathcal{E}}_2)^2
     \right\rangle
    +\frac{1}{2m}\left\langle V \right\rangle
                 \left\langle H_B \right\rangle.
\end{array}
\end{equation}
A respective transformation of the Breit-Pauli operator is
\begin{equation}\label{eq:BPm}
\begin{array}{@{}l}
\displaystyle
H'_B = -\frac{p^4}{8m^3}
   + \frac{\pi}{m^2}[Z_1\delta(\mathbf{r}_1)+Z_2\delta(\mathbf{r}_2)]
\\[2mm]\displaystyle\hspace{30mm}
   -\frac{1}{4m^2}
       (\boldsymbol{\mathcal{E}}_1\!+\!\boldsymbol{\mathcal{E}}_2)
       \boldsymbol{\nabla}
   +2U(H_0-E_0)
\end{array}
\end{equation}
Finally, the energy shift can be written
\begin{equation}\label{a6}
\Delta E^{(6)} =
   \left\langle H'_B Q (E_0-H_0)^{-1} Q H'_B \right\rangle
  +\left\langle H'^{(6)} \right\rangle
  +\left\langle H^{(6)} \right\rangle
\end{equation}
And now divergent terms are gathered together into the modified effective
Hamiltonian $H^{(6)}+H'^{(6)}$.

\subsection{Removing divergences from the modified effective Ha\-miltonian}

The explicit form for the expectation value of the modified effective
Hamiltonian can be written:
\begin{equation}\label{H6m}
\begin{array}{@{}l}
\displaystyle
\Bigl\langle H^{(6)} \Bigr\rangle + \Bigl\langle H'^{(6)} \Bigr\rangle =
   \frac{\left\langle p^6 \right\rangle}{16m^5}
  +\frac{5
      \left\langle
        (\boldsymbol{\mathcal{E}}_1\!+\!\boldsymbol{\mathcal{E}}_2)^2
      \right\rangle}{32m^3}
  +\frac{9\left\langle p^4V\!+\!Vp^4 \right\rangle}{128m^4}
\\[3mm]\displaystyle\hspace{30mm}
  +\frac{\pi\left\langle V(\rho_1\!+\!\rho_2) \right\rangle}{2m^3}
  -\frac{5\left\langle V^3 \right\rangle}{16m^2}
  -\frac{3\pi E_0\left\langle(\rho_1\!+\!\rho_2)\right\rangle}{4m^3}
\\[3mm]\displaystyle\hspace{30mm}
  +\frac{5E_0\left\langle V^2 \right\rangle}{8m^2}
  -\frac{5E_0^2\left\langle V \right\rangle}{16m^2}
  +\frac{\left\langle V \right\rangle\left\langle H_B \right\rangle}{2m}.
\end{array}
\end{equation}
The first five terms are divergent while the remaining part is finite. In
Appendix B it is shown how the divergent terms may be transformed in a
proper way to make them suitable for separation of divergent part, which
then can be cancel out. After summing up of all the terms of the
expression (\ref{H6m}), one gets
\begin{equation}\label{Hmain}
\begin{array}{@{}l}
\displaystyle
\Bigl\langle H^{(6)} \Bigr\rangle + \Bigl\langle H'^{(6)} \Bigr\rangle =
   \frac{3E_0\left\langle V^2 \right\rangle}{4m^2}
  -\frac{5E_0^2\left\langle V \right\rangle}{4m^2}
  -\frac{3\pi E_0\left\langle(\rho_1+\rho_2)\right\rangle}{4m^3}
\\[3mm]\displaystyle\hspace{30mm}
  +\frac{\left\langle\mathbf{p}V^2\mathbf{p}\right\rangle}{8m^3}
  +\frac{\left\langle V \right\rangle\left\langle H_B \right\rangle}{2m}
  +\frac{E_0^3}{2m^2}.
\end{array}
\end{equation}
All the expectation values in (\ref{Hmain}) are finite. Here $E_0(R)$ is
the ground state energy of the two-center problem
(Eq.~(\ref{Hamiltonian})) at a given bond length $R$.

\subsection{Spin-orbit part}

The spin-orbit second order iteration contribution does not contain
divergent part and can be treated separately in a usual way
\begin{equation}\label{eq:so}
\Delta E^{(6)}_{so} =
      \Bigl\langle
         H_B^{so}\>Q\>(E_0-H_0)^{-1}\>Q\>H_B^{so}
      \Bigr\rangle,
\end{equation}
where
\[
H_B^{so}=
      \left(
          Z_1\frac{[\mathbf{r}_1\times\mathbf{p}]}{2m^2r_1^3}+
          Z_2\frac{[\mathbf{r}_2\times\mathbf{p}]}{2m^2r_2^3}
      \right)\mathbf{s}\>,
\]

For $\sigma$-states, $\partial_\varphi\equiv 0$, and without loss of
generality we can consider that our first order solution has a spin state
$s_z=1/2$, then the second order spin-orbit contribution can be rewritten
in terms of variables $r_1$ and $r_2$ as
\begin{equation}\label{eq:so2}
\begin{array}{@{}l}
\displaystyle
\Delta E^{(6)}_{so} =
\biggl\langle
       \frac{irRe^{-i\varphi}}{4m^2}
       \left(
          \frac{Z_1}{r_1^3r_2}\>\partial_{r_2}\!-\!
          \frac{Z_2}{r_1r_2^3}\>\partial_{r_1}
       \right)\!\Psi_0
   \biggm|(E_0\!-\!H_0)^{-1}\biggm|
\\[3mm]\displaystyle\hspace{45mm}
       \frac{irRe^{-i\varphi}}{4m^2}
       \left(
          \frac{Z_1}{r_1^3r_2}\>\partial_{r_2}\!-\!
          \frac{Z_2}{r_1r_2^3}\>\partial_{r_1}
       \right)\!\Psi_0
\biggr\rangle.
\end{array}
\end{equation}

\section{Results and Conclusion}

Equations (\ref{a6}), (\ref{Hmain}), and (\ref{eq:so2}) have been evaluated
numerically using the variational expansion (\ref{exp})--(\ref{expsym}).

In Tables 1 and 2 the relativistic corrections  of order $m\alpha^6$ for
the ground state of electron for symmetric $Z_1\!=\!Z_2\!=\!1$
($\mbox{H}^{+}_{2}$ molecular ion) and asymmetric (the antiprotonic helium
atom) $Z_1\!=\!2$ and $Z_2\!=\!-1$ cases are presented as functions of
a bond length. The accuracy of obtained results for $\Delta E^{(6)}$ is
estimated as all digits indicated in the Tables. Figures 1 and 2
shows the "effective" potentials for these two cases. Dashed lines
are the radial wave function for the ground and first vibrational
$S$-states of $\mbox{H}_2^+$ and the $(36,34)$ state of the
$^4\mbox{He}^+\bar{p}$ atom.

The last table gives a comparison of our calculations with the earlier
ones, which demonstrate superiority of the newly obtained results.
However, we want to draw attention to the approach of \cite{Rut87}, which,
to our opinion, is very promising and allows to evaluate the $m\alpha^8$
order corrections as well using the same first order perturbation wave
function with a rather high accuracy. Presumably, the not very
high precision of this particular calculation is connected with the
Gaussian basis set, which is not very suitable for description of the
zero-order and first-order solutions.

\section*{Acknowledgments}

This work has been supported by the Russian Foundation for Basic Research
under the grant No. 05-02-16618.

\appendix
\section{Analytical evaluation of the matrix elements}

The calculation of the matrix elements is reduced to evaluation of
integrals of the type
\begin{equation}
 \Gamma_{lm}(\alpha,\beta) =
   \int r^{l-1}_{1}r^{m-1}_{2}e^{-\alpha r_{1}-\beta r_{2}}
                                            d^{3}\mathbf{r}.
\end{equation}
Integers $(l, m)$ are, in general, non-negative, but in case of singular
matrix elements one of the indices can be negative.

The function $\Gamma_{00}$ can be easily obtained
\begin{equation}\label{A.2}
\Gamma_{00}(\alpha,\beta,R) = \frac{4\pi}{R}\,
    \frac{e^{-\beta R} - e^{-\alpha R}}{\alpha^{2} - \beta^{2}},
\end{equation}
where $R$ is the distance between nuclei, then
$\Gamma_{lm}(\alpha,\beta;R)$ for non-negative $(l, m)$ may be generated
from (\ref{A.2}) by means of relation
\begin{equation}\label{A.3}
\Gamma_{lm}(\alpha,\beta;R) =
  \left( - \frac{\partial}{\partial \alpha}\right)^{l}
  \left( - \frac{\partial}{\partial \beta}\right)^{m}
  \Gamma_{00}(\alpha,\beta,R).
\end{equation}

Integral $\Gamma_{-1,0}(\alpha,\beta;R)$ is expressed by
\begin{equation}\label{A.4}
\begin{array}{@{}l}
\displaystyle
\Gamma_{-1,0}(\alpha,\beta;R) =
   \frac{2\pi}{R\beta}
      \Bigl\{
         e^{\beta R} \mbox{E}_{1}(R(\alpha+\beta))
         +e^{-\beta R}\ln R(\alpha +\beta)
\\[3mm]\hspace{40mm}
         -e^{\beta R}\bigl[\mbox{E}_{1}(R(\alpha -\beta))
         +\ln R(\alpha -\beta)\bigr]
      \Bigr\}.
\end{array}
\end{equation}
Worthy to note that a function in square brackets is analytic when
argument is zero. Integrals $\Gamma_{-1,m}$ are generated from
$\Gamma_{-1,0}$ similar to (\ref{A.3}):
\begin{equation}\label{A.5}
\Gamma_{-1,m}(\alpha,\beta;R) =
  \left( - \frac{\partial}{\partial \beta}\right)^{m}
  \Gamma_{-1,0}(\alpha,\beta,R).
\end{equation}

Function $\mbox{E}_1(z)$ encountered in (\ref{A.4}) is the exponential
integral function \cite{Abr}:
\[
 \mbox{E}_{1}(z) = \Gamma(0,z) = \int^{\infty}_{z} t^{-1}e^{-t}dt.
\]

\section{Relations between divergent matrix elements}

In this section we will assume that $V$ is regularized in some or other
way and $\rho=\rho_1\!+\!\rho_2$ ($\Delta V\!=\!4\pi\rho$) is a smooth
function of space variables, a distribution of charge in space. The left-
and right-hand side functions inside the brackets represent the same wave
function $\Psi_0$, the solution of the Schr\"odinger equation
(\ref{nonrel}). Then by using commutation relations and integration by
parts one gets
\begin{subequations}
\begin{equation}\label{B1}
\left\langle Vp^2V \right\rangle =
   \left\langle V^2p^2 \right\rangle
   -4\pi\left\langle V(\rho_1\!+\!\rho_2) \right\rangle
   +2\left\langle
      V(\boldsymbol{\mathcal{E}}_1\!+\!\boldsymbol{\mathcal{E}}_2)
                                               \boldsymbol{\nabla}
   \right\rangle,
\end{equation}
\begin{equation}\label{B2}
\left\langle Vp^2V \right\rangle =
   \left\langle
      (\boldsymbol{\mathcal{E}}_1\!+\!\boldsymbol{\mathcal{E}}_2)^2
   \right\rangle
   -2\left\langle
      V(\boldsymbol{\mathcal{E}}_1\!+\!\boldsymbol{\mathcal{E}}_2)
                                               \boldsymbol{\nabla}
   \right\rangle
   +\left\langle \mathbf{p}V^2\mathbf{p} \right\rangle,
\end{equation}
\begin{equation}\label{B3}
\vrule width 0pt height 12pt
4\pi\bigl\langle V(\rho_1\!+\!\rho_2) \bigr\rangle =
   -\left\langle
      (\boldsymbol{\mathcal{E}}_1\!+\!\boldsymbol{\mathcal{E}}_2)^2
   \right\rangle
   +2\left\langle
      V(\boldsymbol{\mathcal{E}}_1\!+\!\boldsymbol{\mathcal{E}}_2)
                                               \boldsymbol{\nabla}
   \right\rangle.
\end{equation}
\end{subequations}
Subtracting Eq.~(\ref{B3}) from Eq.~(\ref{B1}), we have
\begin{subequations}
\begin{equation}\label{B4}
\left\langle Vp^2V \right\rangle =
   \left\langle
      (\boldsymbol{\mathcal{E}}_1\!+\!\boldsymbol{\mathcal{E}}_2)^2
   \right\rangle
   +\left\langle V^2p^2 \right\rangle,
\end{equation}
then summing up (\ref{B2}) and (\ref{B3}) and taking into account
(\ref{B4}):
\begin{equation}\label{B5}
4\pi\bigl\langle V(\rho_1\!+\!\rho_2) \bigr\rangle =
   -\left\langle
      (\boldsymbol{\mathcal{E}}_1\!+\!\boldsymbol{\mathcal{E}}_2)^2
   \right\rangle
   -\left\langle V^2p^2 \right\rangle
   +\left\langle \mathbf{p}V^2\mathbf{p} \right\rangle.
\end{equation}
\end{subequations}

From the last two formulas with a systematic use of
\[
p^2\Psi_0\!=\!2m(E_0\!-\!V)\Psi_0,
\]
the final expressions for the divergent terms in Eq.~(\ref{H6m}) may be
obtained:
\begin{equation}
\begin{cases}
\displaystyle
\frac{\left\langle p^6\right\rangle}{16m^5} =
   \frac{
     \left\langle
       (\boldsymbol{\mathcal{E}}_1\!+\!\boldsymbol{\mathcal{E}}_2)^2
     \right\rangle}{4m^3}
  -\frac{\left\langle V^3\right\rangle}{2m^2}
  +\frac{3E_0\left\langle V^2\right\rangle}{2m^2}
  -\frac{3E_0^2\left\langle V\right\rangle}{2m^2}
  +\frac{E_0^3}{2m^2}\,,\hspace{-5mm}
\\[4mm]\displaystyle
\frac{\left\langle p^4V\!+\!Vp^4 \right\rangle}{128m^4} =
  -\frac{
     \left\langle
       (\boldsymbol{\mathcal{E}}_1\!+\!\boldsymbol{\mathcal{E}}_2)^2
     \right\rangle}{32m^3}
  +\frac{\left\langle V^3\right\rangle}{16m^2}
  -\frac{E_0\left\langle V^2\right\rangle}{8m^2}
  +\frac{E_0^2\left\langle V\right\rangle}{16m^2}\,,
\\[4mm]\displaystyle
\frac{\pi\left\langle V(\rho_1\!+\!\rho_2) \right\rangle}{2m^3} =
  -\frac{
     \left\langle
       (\boldsymbol{\mathcal{E}}_1\!+\!\boldsymbol{\mathcal{E}}_2)^2
     \right\rangle}{8m^3}
  +\frac{\left\langle V^3 \right\rangle}{4m^2}
  -\frac{E_0\left\langle V^2 \right\rangle}{4m^2}
  +\frac{\left\langle\mathbf{p}V^2\mathbf{p}\right\rangle}{8m^3}\,.
  \hspace{-5mm}
\end{cases}
\end{equation}
After summing up all the terms in (\ref{H6m}) the regularization can be
removed.

\clearpage

\begin{figure}
\caption{Adiabatic "effective potential" for the $m\alpha^6$ order
relativistic correction, $\Delta E^{(6)}$ ($Z_1=Z_2=1$) and radial
wave functions of the first two vibrational states ($L=0$) of the
$\mbox{H}_2^+$ molecular ion.}
\begin{center}
\includegraphics*[width=0.6\textwidth]{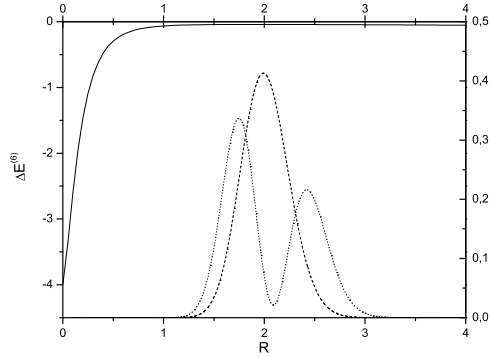}
\end{center}
\end{figure}

\begin{figure}
\caption{Adiabatic "effective potential" for the $m\alpha^6$ order
relativistic correction, $\Delta E^{(6)}$ ($Z_1=2$, $Z_2=-1$) and a radial
wave function of the $(L=34,v=2)$) state of the $^4\mbox{He}^+\bar{p}$
atom.}
\begin{center}
\includegraphics*[width=0.6\textwidth]{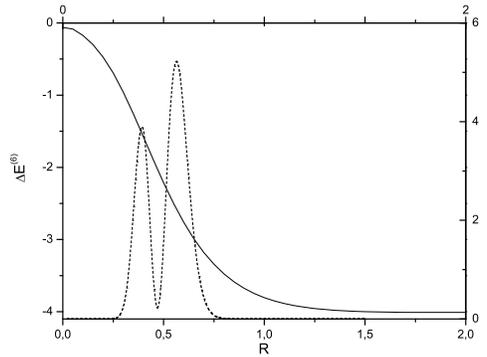}
\end{center}
\end{figure}

\begin{table}
\caption{Adiabatic potentials of the nonrelativistic energy and the
$m\alpha^6$ order relativistic correction, $\Delta E^{(6)}$, for $Z_1=Z_2=1$.}
\begin{tabular}{c@{\hspace{4mm}}c@{\hspace{4mm}}c@{\hspace{7mm}}
                c@{\hspace{4mm}}c@{\hspace{4mm}}c}
\hline\hline
 $R$ & $E_{nr}$ & $\Delta E^{(6)}$ & $R$ & $E_{nr}$ & $\Delta E^{(6)}$ \\
\hline
 0.00 & $-$2.0000000000 & $-$4.000000 &  2.60 & $-$0.9754485809 & $-$0.043808 \\
 0.05 & $-$1.9939766092 & $-$3.367825 &  2.70 & $-$0.9580276600 & $-$0.044356 \\
 0.10 & $-$1.9782420708 & $-$2.607169 &  2.80 & $-$0.9414988606 & $-$0.044941 \\
 0.15 & $-$1.9557215572 & $-$1.960310 &  2.90 & $-$0.9258056314 & $-$0.045558 \\
 0.20 & $-$1.9286203017 & $-$1.462814 &  3.00 & $-$0.9108961973 & $-$0.046204 \\
 0.25 & $-$1.8985571104 & $-$1.093674 &  3.10 & $-$0.8967230608 & $-$0.046873 \\
 0.30 & $-$1.8667040794 & $-$0.823325 &  3.20 & $-$0.8832425598 & $-$0.047563 \\
 0.40 & $-$1.8007540594 & $-$0.481321 &  3.30 & $-$0.8704144746 & $-$0.048269 \\
 0.50 & $-$1.7349879999 & $-$0.296145 &  3.40 & $-$0.8582016779 & $-$0.048987 \\
 0.60 & $-$1.6714847144 & $-$0.193143 &  3.50 & $-$0.8465698245 & $-$0.049714 \\
 0.70 & $-$1.6111962656 & $-$0.134093 &  3.60 & $-$0.8354870739 & $-$0.050447 \\
 0.80 & $-$1.5544800944 & $-$0.099226 &  3.70 & $-$0.8249238441 & $-$0.051181 \\
 0.90 & $-$1.5013815992 & $-$0.078065 &  3.80 & $-$0.8148525916 & $-$0.051913 \\
 1.00 & $-$1.4517863133 & $-$0.064901 &  3.90 & $-$0.8052476157 & $-$0.052640 \\
 1.10 & $-$1.4055027761 & $-$0.056534 &  4.00 & $-$0.7960848837 & $-$0.053358 \\
 1.20 & $-$1.3623078578 & $-$0.051122 &  4.20 & $-$0.7789974438 & $-$0.054756 \\
 1.30 & $-$1.3219713911 & $-$0.047577 &  4.40 & $-$0.7634258673 & $-$0.056082 \\
 1.40 & $-$1.2842692423 & $-$0.045241 &  4.60 & $-$0.7492240944 & $-$0.057317 \\
 1.50 & $-$1.2489898721 & $-$0.043709 &  4.80 & $-$0.7362614458 & $-$0.058443 \\
 1.60 & $-$1.2159372245 & $-$0.042725 &  5.00 & $-$0.7244202951 & $-$0.059450 \\
 1.70 & $-$1.1849315636 & $-$0.042124 &  5.20 & $-$0.7135942526 & $-$0.060330 \\
 1.80 & $-$1.1558091896 & $-$0.041799 &  5.40 & $-$0.7036867525 & $-$0.061083 \\
 1.90 & $-$1.1284215723 & $-$0.041678 &  5.60 & $-$0.6946099538 & $-$0.061711 \\
 2.00 & $-$1.1026342144 & $-$0.041711 &  5.80 & $-$0.6862838773 & $-$0.062222 \\
 2.10 & $-$1.0783254220 & $-$0.041866 &  6.00 & $-$0.6786357151 & $-$0.062624 \\
 2.20 & $-$1.0553850811 & $-$0.042118 &  7.00 & $-$0.6484511470 & $-$0.063445 \\
 2.30 & $-$1.0337134948 & $-$0.042449 &  8.00 & $-$0.6275703886 & $-$0.063298 \\
 2.40 & $-$1.0132203052 & $-$0.042848 &  9.00 & $-$0.6123065640 & $-$0.062991 \\
 2.50 & $-$0.9938235109 & $-$0.043304 & 10.00 & $-$0.6005787289 & $-$0.062766 \\
\hline\hline
\end{tabular}
\end{table}

\begin{table}
\caption{Adiabatic potentials of the nonrelativistic energy and the
$m\alpha^6$ order relativistic correction, $\Delta E^{(6)}$, for
$Z_1=2$, $Z_2=-1$.}
\begin{tabular}{c@{\hspace{4mm}}c@{\hspace{4mm}}c@{\hspace{7mm}}
                c@{\hspace{4mm}}c@{\hspace{4mm}}c}
\hline\hline
 $R$ & $E_{nr}$ & $\Delta E^{(6)}$ & $R$ & $E_{nr}$ & $\Delta E^{(6)}$ \\
\hline
 0.00 & $-$0.5000000000 &  $-$0.062500 & 0.90 & $-$1.0519803013 & $-$3.675683 \\
 0.05 & $-$0.5031900327 &  $-$0.083184 & 0.95 & $-$1.0854533401 & $-$3.747452 \\
 0.10 & $-$0.5123196419 &  $-$0.171722 & 1.00 & $-$1.1174174054 & $-$3.804849 \\
 0.15 & $-$0.5269067301 &  $-$0.298874 & 1.10 & $-$1.1768579850 & $-$3.886695 \\
 0.20 & $-$0.5465679498 &  $-$0.472257 & 1.20 & $-$1.2305595991 & $-$3.937659 \\
 0.25 & $-$0.5709081980 &  $-$0.693842 & 1.30 & $-$1.2789566171 & $-$3.968875 \\
 0.30 & $-$0.5994601342 &  $-$0.959464 & 1.40 & $-$1.3225517671 & $-$3.987627 \\
 0.35 & $-$0.6316629543 &  $-$1.258908 & 1.50 & $-$1.3618542175 & $-$3.998586 \\
 0.40 & $-$0.6668734353 &  $-$1.577696 & 1.60 & $-$1.3973469293 & $-$4.004719 \\
 0.45 & $-$0.7043992306 &  $-$1.899900 & 1.70 & $-$1.4294714947 & $-$4.007894 \\
 0.50 & $-$0.7435419103 &  $-$2.210995 & 1.80 & $-$1.4586229693 & $-$4.009286 \\
 0.55 & $-$0.7836381742 &  $-$2.499767 & 1.90 & $-$1.4851501332 & $-$4.009626 \\
 0.60 & $-$0.8240915228 &  $-$2.759094 & 2.00 & $-$1.5093584825 & $-$4.009370 \\
 0.65 & $-$0.8643913616 &  $-$2.985702 & 2.50 & $-$1.6038198680 & $-$4.005852 \\
 0.70 & $-$0.9041202195 &  $-$3.179363 & 3.00 & $-$1.6684906434 & $-$4.003217 \\
 0.75 & $-$0.9429517761 &  $-$3.341928 & 4.00 & $-$1.7505678833 & $-$4.001067 \\
 0.80 & $-$0.9806429215 &  $-$3.476445 & 5.00 & $-$1.8002303298 & $-$4.000427 \\
 0.85 & $-$1.0170227063 &  $-$3.586486 & 6.00 & $-$1.8334437319 & $-$4.000196 \\
\hline\hline
\end{tabular}
\end{table}

\begin{table}
\caption{Comparison with earlier calculations at a bond length $R=2.0$ [Bohr].}
\begin{tabular}{@{}c@{\hspace{6mm}}l@{\hspace{6mm}}l@{\hspace{6mm}}l@{}}
\hline\hline
 & \multicolumn{1}{c}{$E^{(0)}$~~~~} & \multicolumn{1}{c}{$E^{(1)}c^2$~~~~}
 & \multicolumn{1}{c}{$E^{(2)}c^4$} \\
\hline
this work
& $-1.102\>634\>214\>494\>946\>461\>50$ & $-0.138\>332\>9939$ & $-0.041\>711$ \\
\cite{Mark87}
& $-1.102\>634\>206$ & $-0.138\>325$ & $-$0.0417 \\
\cite{Rut87}
& $-1.102\>600$ & $-0.138\>277$ & $-$0.0399 \\
\cite{How90}
&               & $-0.138\>333$ &           \\
\hline\hline
\end{tabular}
\end{table}

\end{document}